\documentclass[twocolumn,twoside,english]{article}

\usepackage{graphicx}

\begin{document}

\thispagestyle{plain}
\markboth{\hfill \rm \uppercase{On Particles Creation and Renormalization}
 \hfill}{\hfill \rm \uppercase{Pavlov} \hfill }

\twocolumn[
\begin{center}
{\LARGE \bf On Particles Creation and Renormalization

\vspace{10pt}
in a Cosmological Model with a Big Rip}
\vspace{12pt}

{\Large {Yu. V. Pavlov}${}^*$} \vspace{12pt}

{\it Institute of Mechanical Engineering, Russian Acad. Sci., %\\
Bol'shoy pr. 61, St. Petersburg 199178, Russia;}
\vspace{4pt}

{\it A. Friedmann Laboratory for Theoretical Physics, %\\
Griboedov kanal 30/32, St.\,Petersburg 191023, Russia}
\end{center}
\vspace{5pt}
 {\bf Abstract.}
    An exact solution is obtained for a massive scalar field conformally
coupled with the curvature in a cosmological model with the scale
factor $a(t) = a_0 / |t|$, corresponding to background matter with the
equation of state $ p=- 5 \varepsilon /3 $.
    An expression for the number density of created particles is obtained,
and its behavior is studies as the model approached the instant of a Big Rip.
    Renormalization of the energy-momentum tensor is considered,
and it is shown that back reaction of the quantum effects of a conformally
coupled scalar field on the space-time metric can be neglected if
the field mass is much smaller than the Planck mass and if the time
left to the Big Rip is greater than the Planck time.

\vspace{10pt}
{PACS number:} {\tt 04.62.+v}
\vspace{27pt}
]

%%%%******************************************************************
{\centering  \section{Introduction}}

\footnotetext[1]{E-mail:\, {\tt yuri.pavlov@mail.ru}}

\vspace{-4pt}
    Particle creation in curved space-time has been actively studied since
the 70s of last century.
    Results were obtained which may have important applications in cosmology
and astrophysics (see~\cite{GMM}).
    In particular, creation of particles with mass on the Grand Unification
scale by the gravitational field of the early Universe may be used for
explaining the observed density of visible and dark matter~\cite{GrPv} and
be related to observations of superhigh-energy cosmic rays~\cite{GrPvAGN}.

   In connection with the recent discovery of the accelerated expansion of
the Universe, it is of interest to study quantum effects in the cosmological
background determined by matter with negative pressure, in particular,
phantom matter.
    In this paper, we consider particle creation in cosmological model that
admits an exact solution of the scalar field equation.
    This exact solution makes it possible to analyze the behavior of
expressions for the number density of created particles as the cosmological
time tends to the instant of a Big Rip.
    We also consider renormalization of the energy-momentum tensor~(EMT)
and study the back reaction of quantum effects of the scalar field
on the space-time metric.

    We use the system of units in which $\hbar = c \!=\! 1$. \
    The signs of the Riemann and Ricci tensors are chosen in such a way that
$ R^{\, i}_{\ jkl} = \partial_l \, \Gamma^{\, i}_{\, jk} -
\partial_k \, \Gamma^{\, i}_{\, jl} +
\Gamma^{\, i}_{\, nl} \Gamma^{\, n}_{\, jk} -
\Gamma^{\, i}_{\, nk} \Gamma^{\, n}_{\, jl}  $,
$\ R_{ik} = R^{\, l}_{\ ilk}$,
where $\Gamma^{\, i}_{\, jk}$ are the Christoffel symbols.

\vspace{11pt}
%%%%*****************************************************************
{%\centering
 \section{Scalar field in a homogeneous isotropic space}
\label{sec2}
}

    Consider a complex scalar field $\varphi(x)$ of mass $m$
with the Lagrangian
    \begin{equation}
L(x)=\sqrt{|g|} \left[\, g^{ik}\partial_i\varphi^*\partial_k\varphi -
(m^2 + \xi_c R)\, \varphi^* \varphi \, \right]
\label{Lag}
\end{equation}
     and the corresponding equation of motion
\begin{equation}
( \nabla^i \nabla_{\! i} + m^2 + \xi_c R)\, \varphi(x)=0 \,,
\label{Eqm}
\end{equation}
    where ${\nabla}_{\! i}$ are covariant derivatives
in $N$-dimensional space-time with the metric~$g_{ik}$,
$R$ is the scalar curvature, $\xi_c = (N-2)/\,[\,4\,(N-1)] $
\ ($\xi_c = 1/6 $ for $N=4$), \ $ g\!=\!{\rm det}(g_{ik})$.\,
    Eq.~(\ref{Eqm}) is conformally invariant if~$m=0$.

    For homogeneous isotropic space-time with the metric
    \begin{equation}
ds^2= d t^2- a^2(t)\, d l^2 = a^2(\eta)\,(d{\eta}^2 - d l^2) \,,
\label{gik}
\end{equation}
    where $d l^2 $ is the metric of $(N-1)$-dimensional space
with constant curvature $K=0, \pm 1 $,
the full set of solutions to Eq.~(\ref{Eqm}) can be found in the form
    \begin{equation}
\varphi(x) = a^{-(N-2)/2} (\eta)\, g_\lambda (\eta) \Phi_J ({\bf x}) \,,
\label{fgf}
\end{equation}
        where
    \begin{equation}
g_\lambda''(\eta)+\omega^2(\eta)\,g_\lambda(\eta)=0 \,,
\label{gdd}
\end{equation}
       \begin{equation}
\omega^2(\eta)= m^2 a^2(\eta) +\lambda^2 \,,
\label{Ome}
\end{equation}
     \begin{equation}
\Delta_{N-1}\,\Phi_J ({\bf x}) = - \Biggl( \lambda^2 -
\biggl( \frac{N-2}{2} \biggr)^2 K \Biggr) \Phi_J  ({\bf x})\,,
\label{DFlF}
\end{equation}
    $J$ is the set of indices (quantum numbers) enumerating the eigenfunctions
of the Laplace-Beltrami operator~$\Delta_{N-1}$ in
($N-1$)-dimensional space.

    In accordance with the Hamiltonian diagonalization method~\cite{GMM},
the functions~$g_\lambda(\eta)$ should satisfy the following initial
conditions~\cite{Pv2001}:
    \begin{equation}
g_\lambda'(\eta_0)=i\, \omega(\eta_0)\, g_\lambda(\eta_0) \,, \ \
|g_\lambda(\eta_0)|= \omega^{-1/2}(\eta_0).
\label{icg}
\end{equation}

    If the quantized scalar field is in a vacuum state for the time
instant~$\eta_0$, then the number density of particle pairs created by
the instant~$\eta $, can be calculated
(for the quasi-Euclidean metric with~$K=0$) by the formula~\cite{GMM}
    \begin{equation}
n(\eta) = \frac{B_N}{2 a^{N-1}} \int \limits_0^\infty \! S_\lambda(\eta)\,
\lambda^{N-2}\, d \lambda,
\label{nN}
\end{equation}
    where
$B_N=\left[2^{N\!-3} \pi^{(N \!-1)/2} \Gamma((N\!-1)/2) \right]^{-1}\!,$
\ $\Gamma(z)$ is the gamma function, and
    \begin{equation}
S_\lambda(\eta) = \left| g'_\lambda (\eta ) -
i \omega \, g_\lambda (\eta ) \right|^2 / \, (4 \omega ) \,.
\label{Sgg}
\end{equation}
    As has been shown in~\cite{Pv2001}, $ S_\lambda \sim \lambda^{-6} $,
and the integral in~(\ref{nN}) converges for~$N<7$.

\vspace{14pt}
%%%%*****************************************************************
{\centering \section{A cosmological model with phantom matter}
\label{sec3}
}

    Consider a cosmological model with the background mater having
the equation of state $ p =w \varepsilon $, where $ w < -1 $.
    from the Einstein equations
    \begin{equation}
R_{ik} - \frac{1}{2} g_{ik} R = - 8 \pi G T_{ik} \,,
\label{Ein}
\end{equation}
    where $ T^i_k = {\rm diag}\, ( \varepsilon, -p, \ldots , -p\,) $,
it follows that, in the metric~(\ref{gik}), the energy density of
the background matter grows according to the law
    \begin{equation}
\varepsilon (\eta) \sim a^{- (1+w)(N-1)}(\eta) \,.
\label{ep}
\end{equation}
    For $K=0$, from~(\ref{Ein}) we obtain
    \begin{equation}
a = a_0 / (-t)^{q} = a_1 / (-\eta)^{\beta} \,,
\label{ate}
\end{equation}
    where $ t \in (- \infty, 0), \ \ \eta \in (- \infty, 0)$,
    \begin{equation}
\beta = - \frac{2}{N \!-3 + w (N \!-1)^{\mathstrut} } \ , \ \ \
q = \frac{\beta}{1 - \beta^{\mathstrut} } \,.
\label{beq}
\end{equation}
    For $w<-1$, $\beta \in (0,1) $, and there is a ``Big rip''
singularity~\cite{Starobinsky00} at $t \to -0$.

%%%%%%%%%%%%%%%%%%%%%%%%%%%%%%%%%%%%%%%%%

    For the value $w=-(N+1)/(N-1)$
(so that $ w = -5/3 $  in four-dimensional space-time),
when $ q = 1 \,, \ \beta = 1/2 $, \, Eq.~(\ref{gdd})
can be solved exactly in terms of degenerate hypergeometric functions
(see (2.1.2.103) in Ref.~\cite{ZaiPol01}).
    The solution satisfying the condition~(\ref{icg}) as $\eta_0 \to -\infty$,
has the form
    \begin{eqnarray}
g_\lambda(\eta)= -2i \eta \sqrt{\lambda}\, \exp \Bigl(\!
-\frac{\pi m^2 a_1^2}{4 \lambda} + i ( \lambda \eta +\alpha_0)\! \Bigr)
\times \hspace{-4mm} \nonumber \\ \times \,
\Psi \Bigl(\! 1+ \frac{i m^2 a_1^2}{2 \lambda}\, , 2\, ;
-2 i \lambda \eta \! \Bigr), \hspace{4mm}
\label{solBR}
\end{eqnarray}
    where $ \Psi ( a ,\, b ;\, z ) $ is Tricomi's degenerate hypergeo\-metric
function, and~$\alpha_0 $ is an arbitrary real constant.
    In what follows, we will suppose that quantized scalar field is in a vacuum
state for a time instant $ \eta_0 \to -\infty $.

%%%%%%%%%%%%%%%%%%%%%%%%%%%%%%%%%%%%%%%%%%%%%%%%%%
    Let us introduce $p_\lambda=\lambda/(ma)$,
the physical momentum $\lambda / a $ measured in the units of~$m$.
    Making the corresponding substitution in~(\ref{nN}), we obtain for
the density of particles created by the instant~$t$:
    \begin{equation}
n(t) =  m^{N-1} \frac{B_N}{2} \int \limits_0^\infty \!
S_{p_\lambda}(t)\, p_\lambda^{N-2} d p_\lambda,
\label{nNp}
\end{equation}
    where, in agreement with~(\ref{Sgg}) and~(\ref{solBR})
    \begin{eqnarray}
S_{p_\lambda}(t) = \frac{p_\lambda \, e^{\pi m t/(4 p_\lambda)}}
{16 \sqrt{\mathstrut 1+p_\lambda^2} }
\left| \Bigl[ \Bigl( \sqrt{1+p_\lambda^2}- p_\lambda \Bigr)   \right. \times
\nonumber \\
\times\,  i 2 m t - 4 \Bigr]
\Psi \Bigl( 1 \!- \frac{i m t}{4 p_\lambda}\, ,\, 2\, ; - i p_\lambda m t
\Bigr) - \nonumber \\
- \left. m t (m t + i 4 p_\lambda)
\Psi \Bigl( 2 \!- \frac{i m t}{4 p_\lambda}\, ,\, 3\, ;
- i p_\lambda m t \Bigr) \right|^{\,2}\!\!\!\!.
\label{Sp}
\end{eqnarray}

    Let us find the asymptotic $S_{p_\lambda}(t) $ for $t \to -0$.
From Eq. 6.7.(13) from~\cite{BatemanErdelyi} we obtain, as $z \to 0$,
    \begin{equation}
\Psi \bigl( 1 + \alpha z,\, 2 ;\, z \bigr) =
\frac{1}{z\, \Gamma(1 + \alpha z)} + O(|z \ln z|) \,,
\label{as1}
\end{equation}
    \begin{equation}
\Psi \bigl( 2 + \alpha z,\, 3 ;\, z \bigr) =
\frac{1}{\Gamma(2 \!+ \alpha z)} \!
\Bigl( \! \frac{1}{z^2} - \alpha \!\Bigr) + O(|z \ln z|).
\label{as2}
\end{equation}
    Therefore,
    \begin{equation}
S_{p_\lambda}(t) \sim \frac{\Bigl(\sqrt{1+ p_\lambda^{\mathstrut 2}}
- p_\lambda \Bigr)^2}{4 p_\lambda\,
\sqrt{ 1+ p_\lambda^{\mathstrut 2}}} \,, \ \ \ \ \ t \to - 0 \,.
\label{asSp}
\end{equation}
    Let us note that this an asymptotic as $ t \to - 0 $.
    The main term of the asymptotic $ S_{p_\lambda}(t) $
as $ p_\lambda \to \infty $ is proportional to $ p_\lambda^{-6} $
and has the form $ S_{p_\lambda}(t) \sim 1/ (16 p_\lambda^6 m^2 t^2)$.

\begin{figure}[ht]
\centering
\includegraphics[width=77mm]{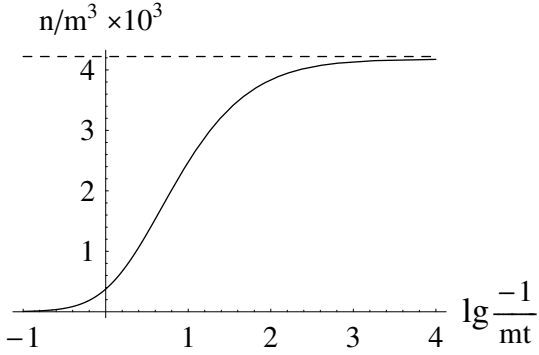}\\[-7pt]
\caption{Time dependence of the number density of created particles
($w=-5/3$).}
\label{BR1}
\end{figure}

    Fig.~\ref{BR1} shows the time dependence of the number density of created
particles for~$N=4$.
%%%%%%%%%%%%%%%%%%%%%%%%%%%%%%%%%%%%%%%%%%%%%%%%%

    The dashed line corresponds to the asymptotic value
$ n = m^3 / (24 \pi^2) $ which can be obtained by integration~(\ref{asSp})
in the exprassion~(\ref{nNp}).
    As is seen from Fig.~\ref{BR1}, particle creation mostly begins near
the Compton time~$t_C=-1/m $ before the Big Rip.

\vspace{14pt}
%%%%*****************************************************************
{\centering \section{Renormalization of the energy-momentum tensor}
\label{sec4}
}

    In obtaining the renormalized vacuum EMT, let us note that,
in the homogeneous, isotropic space with~$K=0$, the vacuum EMT is diagonal,
$ \langle T_i^k \rangle_{\rm ren} = {\rm diag}
(\varepsilon_\varphi, -p_\varphi, \ldots, -p_\varphi)$,
and its components satisfy the covariant conservation condition
    \begin{equation}
(\varepsilon_\varphi)' + \frac{a'(\eta)}{a(\eta)} (N-1)
(\varepsilon_\varphi + p_\varphi) = 0.
\label{epren}
\end{equation}

    For a scalar field conformally coupled to the curvature,
the vacuum EMT for~$N=4$ and~$K=0$ can be presented in the form
    \begin{equation}
\langle T_i^k \rangle_{\rm ren} = \tilde{T}_i^k +
\frac{m^2 G_i^k}{144 \pi^2} +
\frac{1}{1440 \pi^2}\Bigl({}^{(3)}\!H_i^k -
\frac{1}{6}{}^{(1)}\!H_i^k \!\Bigr),
\label{TIKK}
\end{equation}
    where $G_i^k = R_i^k - R g_i^k / 2$ is the Einstein tensor,
the expressions for the tensors ${}^{(1)}\!H_i^k $ and ${}^{(3)}\!H_i^k $
are given, for example, in~\cite{GMM},
    \begin{equation}
\tilde{T}_0^0 \equiv \varepsilon_s
= \frac{1}{\pi^2 a^{4}(\eta)} \int \limits_0^\infty
\! \omega(\eta) S_\lambda (\eta) \lambda^{2}\, d \lambda,
\label{tik1}
\end{equation}
    and
$ \tilde{T}_\alpha^\beta = - \delta_\alpha^\beta p_s $
can be found from~(\ref{epren}), $\alpha , \beta = 1, 2, 3$.
    In the $N$-dimensional case, renormalization of the EMT with the aid
of the Zel'dovich-Starobinsky $n$-wave procedure~\cite{ZlSt}
has been considered in~\cite{Pavlov2004}.

\begin{figure}[t]
\centering
\includegraphics[width=77mm]{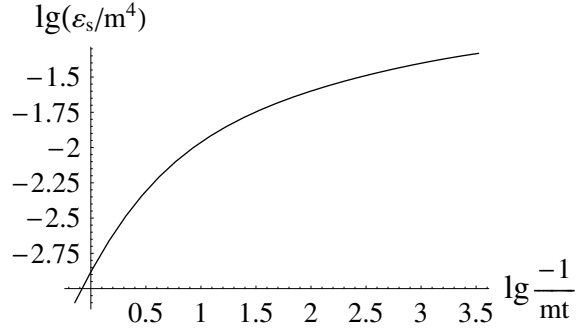}\\[-7pt]
\caption{Time dependence of $\varepsilon_s$.}
\label{EBR}
\end{figure}

   For $a(t)=a_0/(-t)$, from~(\ref{TIKK}) we obtain
    \begin{equation}
\varepsilon_\varphi = \varepsilon_s (t)-
\frac{m^2}{48 \pi^2 t^2} +
\frac{1}{48 \pi^2 t^4},
\label{epe}
\end{equation}
    \begin{equation}
p_\varphi = p_s (t) +
\frac{5 m^2}{144 \pi^2 t^2} -
\frac{7}{144 \pi^2 t^4},
\label{epp}
\end{equation}
    where
    \begin{equation}
\varepsilon_s(t) =  \frac{m^4}{\pi^2} \int \limits_0^\infty \!
S_{p_\lambda}(t)\sqrt{1+p_\lambda^2} \, p_\lambda^{2} \, d p_\lambda.
\label{eee}
\end{equation}
    The calculated results $ \varepsilon_s(t) $ with $ S_{p_\lambda}(t)$
from~(\ref{Sp}) are presented in Fig.~\ref{EBR}.
%%%%%%%%%%%%%%%%%%%%%%%%%%%%%%%%%%%%%%%%%%%%%%%%%

    As is seen from Fig.~\ref{EBR}, as $t \to -0$, vacuum EMT
is dominated by the second and third terms.
    Let us estimate their influence on the space-time metric.

    according to the Einstein equations, the background matter EMT is
$T_i^k = -M_{Pl}^{\, 2}\, G_i^k /(8 \pi)$ \
(where $M_{Pl} =1 / \sqrt{G} $ is the Planck mass).
    Consequently, the ratio of the second term in~(\ref{TIKK})
to the background EMT is
    \begin{equation}
\frac{\Delta T_i^k(2)}{T_i^k} =
\frac{-1}{18 \pi} \left( \frac{m}{M_{Pl}} \right)^2
\label{ot2}
\end{equation}
    which is negligibly small for $m \ll M_{Pl}$.
    Note that the conclusion that it is possible to neglect the back reaction
of created particles on the space-time metric in models with phantom matter
was made in~\cite{Fabris} in the case of a massless minimally coupled scalar
field.

    For the third term in the vacuum EMT~(\ref{TIKK}), we obtain
    \begin{equation}
\frac{\Delta T_0^0(3)}{T_0^0} = \frac{1}{18 \pi (t M_{Pl})^2}, \ \
\frac{\Delta T_\alpha^\beta(3)}{T_\alpha^\beta} =
\frac{7}{90 \pi (t M_{Pl})^2}.
\label{ot3}
\end{equation}
    Therefore, the influence of the third polarization term in the vacuum EMT
on the space-time metric becomes important only at times smaller than
Planckian, $|t|<t_{Pl}=1/M_{Pl}$, before the Big Rip.
    At such times, it would also be necessary to take into account not only
the back reaction of a quantum field on the metric~\cite{NojiriOdintsov04}
but also effects of quantum gravity itself.

    We conclude that the back reaction of quantum effects of a massive,
conformally coupled scalar field on the space-time metric, in
the cosmological model under consideration, with an exact solution of
the field equations, can be neglected in the whole region where one can
apply the approach of quantum field theory in curved space-time.

\vspace{11pt}
{\centering \section*{Acknowledgments}}

    The author thanks Prof. A.A.\,Grib and the participants of the seminar
at the A.A.\,Friedmann Laboratory of Theoretical Physics for a discussion
of this work.

\vspace{11pt}
%%%%%%%%%%%%%%%%%%%%%%%%%%%%%%%%%%%%%%%%%%%%%%%%%%%%%%%%%%%%%%%%%%%%%%

\end{document}